\newcommand{\eql}{ < \kern -12pt \lower 5pt \hbox{$\displaystyle =$}}
\newcommand{\eqg}{ > \kern -12pt \lower 5pt \hbox{$\displaystyle =$}}
\newcommand{\lsim}{ < \kern -11.2pt \lower 4.3pt \hbox{$\displaystyle \sim$}}
\newcommand{\gsim}{ > \kern -12pt   \lower 5pt   \hbox{$\displaystyle
    \sim$}}
 
\documentstyle[prl,aps,epsf,12pt]{revtex}
\begin{document}

\title{ \mbox{$\pi$ excitation} of the $t$-$J$ model}

\author{ Eugene Demler$^{1}$,  Hiroshi Kohno$^{2}$,
    and Shou-Cheng Zhang$^{1}$ }
\address{$^{1}$Department of Physics, Stanford University, Stanford,
  CA~~94305 \\
$^2$ Department of Physics, University of Tokyo, Bunkyo-ku, Tokyo 113,
  Japan 
}
\date{\today}
 
\maketitle

\begin{abstract} In this paper, we present analytical and numerical   
  calculations of the $\pi$ resonance in the $t$-$J$ model. 
   We show in detail how the $\pi$ resonance in the particle-particle 
  channel couples to and appears in the dynamical spin correlation 
  function in a superconducting state. 
   The contribution of the $\pi$ resonance to the spin excitation 
  spectrum can be estimated from general model-independent sum rules, 
  and it agrees with our detailed calculations. 
   The results are in overall agreement with the exact diagonalization 
  studies of the $t$-$J$ model. 
   Earlier calculations predicted the correct doping dependence of the 
  neutron resonance peak in the $YBCO$ superconductor, and in this paper 
  detailed energy and momentum dependence of the spin correlation function 
  is presented. 
   The microscopic equations of motion obtained within current
  formalism agree with that of the $SO(5)$ nonlinear sigma model,
  where the $\pi$ resonance is interpreted as a pseudo Goldstone mode
  of the spontaneous $SO(5)$ symmetry breaking. 
\end{abstract}

\section{Introduction}
\label{intro}

Of many fascinating experiments on high $T_c$ superconductors, the 
resonant neutron scattering peak observed in the $YBCO$ family is
an extremely striking one\cite{fong1,fong3,dai1,fong2,dai2}. 
 It was first observed in the optimally doped $YBCO$ materials. 
 The mode exists only in a narrow region in reciprocal 
space near $(\pi/a,\pi/b,\pi/c)$, where $a$ and $b$ are the lattice 
constants in the $CuO_2$ plane and $c$ is the distance between two
neighboring $CuO_2$ planes in a unit cell. 
 (In the following, we will set these lattice constants to unity to 
simplify notations).
 The energy of the resonance is $41meV$ and it disperses weakly in
reciprocal space. 
 Perhaps the most striking property of this mode is its disappearance
above $T_c$. 
 More recently, this type of collective mode has also been observed in 
the underdoped families of the $YBCO$ superconductors. 
 Here the energy of this mode is $33meV$ and $25meV$, for materials with 
$T_c$ values of $62K$ and $52K$ respectively.
 While the mode energy decreases monotonically with $T_c$, the mode 
intensity increases as $T_c$ decreases.
 Compared with the $41meV$ peak, these modes also have a broader 
spectral distribution below $T_c$. 
 In these underdoped materials the resonance is also observed above 
$T_c$ where it becomes significantly broader. 
 All the modes have been observed in the neutron spin flip channel, 
and more recently, the $41meV$ mode was seen to broaden under a 
uniform magnetic field\cite{bourges}, both indicating that the modes 
are spin triplets.

 These striking resonances have generated wide theoretical interests
and a number of theoretical ideas have been suggested in order to
explain their 
properties\cite{PRL,nejat,mazin,liu,onufrieva,blumberg,zha,millis,yin,abrikosov}.
 We believe that one key ingredient is the coupling of the neutron to 
the particle-particle (p-p) channel which occurs in the superconducting (SC) 
state via the condensate. 
 In particular for a $d_{x^2-y^2}$ gap, the coherence factor 
$[ 1 - \Delta_{k+q}\Delta_k/ E_k E_{k+q} ]/2$ for 
$ Q=(\pi,\pi,\pi)$ goes to unity at threshold rather than vanishing as it 
would for an $s$-wave superconductor\cite{fong1,scalapino}. 
 Furthermore, two of us argued that the p-p interactions in this channel 
leads to a sharp resonance which was called the $\pi$-mode\cite{PRL}. 
 In the normal state, the resonance is decoupled from the neutron scattering, 
but can in principle be observed in pair tunneling experiments\cite{bazaliy}. 
 This theory predicted the doping dependence of the mode energy and 
intensity which was subsequently verified experimentally\cite{keimer}. 
 This picture was also later verified in detailed numerical 
calculations of the Hubbard and the $t$-$J$ models\cite{meixner,eder}.

 In this paper, we study the $\pi$ resonance using a self-consistent linear 
%%%response (SCLR)
response 
theory which formally takes into account the mixing of the
particle-hole (p-h) with the p-p channels in the SC state. 
%%% We first overview some general features of the $\pi$-resonance in 
%%%Sec.\ref{general}. 
%%% The SCLR 
 This formalism is explained in Sec.\ref{CF}. 
 In Sec.\ref{numerix}, we present numerical results based on this 
formalism and show the overall structure of the spin correlation function. 
 We then give an approximate but analytic expression for the resonance 
in Sec.\ref{analyt}. 
 In Sec.\ref{EOM}, we compare our formalism with the results obtained by 
using equations of motion for the $t$-$J$ model and with the $SO(5)$ 
quantum non-linear sigma model. 
 In Sec.\ref{summary}, we summarize the results and conclude the paper 
with some general remarks. 
 Before going into these details, we give here some general features of 
the $\pi$-resonance.

%%%\section{General Features of $\pi$ Resonance}
%%%\label{general}

\begin{figure*}[h]
\centerline{\epsfxsize=7cm
\epsfbox{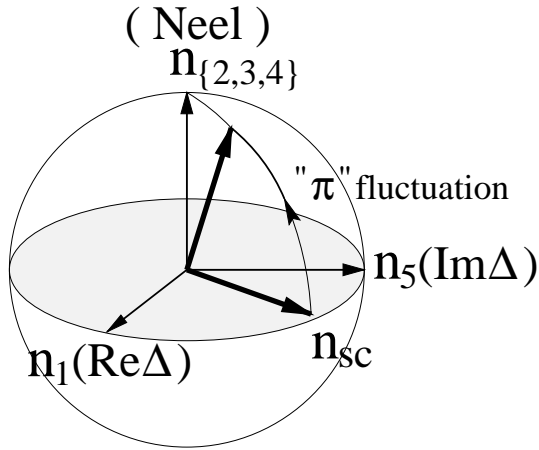}
}
\caption{ Geometric interpretation of the $\pi$ resonance in the
  superspin phase.  }
\label{pi-sphere.eps}
\end{figure*}

 The central object of the theory of the $\pi$ resonance is the so
called $\pi$ operator\cite{PRL}, defined by
\begin{eqnarray}
\pi_{\alpha}^{\dagger}=  \frac{1}{2\sqrt{2}}\sum_p ( cos p_x - cos p_y
  ) c_{p+Q}^{\dagger} 
 \sigma^{\alpha} \sigma^{y} c_{-p}^{\dagger} 
\end{eqnarray}
with $\sigma^{\alpha}$ being Pauli matrices and
$c^{\dagger}_p=(c^{\dagger}_{p\uparrow},c^{\dagger}_{p\downarrow})$ .
 This operator is a spin triplet and carries charge two.
 This operator inspired one of us (SCZ) to formulate the $SO(5)$ theory
of high $T_c$ superconductivity\cite{so5}.
 Together with the total spin and total charge operators, the six $\pi$ 
operators form an $SO(5)$ Lie algebra. 
 A natural vector representation of this $SO(5)$ Lie algebra is the superspin 
\begin{eqnarray}
\vec n &=& (n_1, n_2, n_3, n_4, n_5) \nonumber\\
       &=& (\frac{\Delta + \Delta^{\dagger}}{2}, N_x, N_y, N_z, 
            \frac{\Delta - \Delta^{\dagger} }{2i}) 
\end{eqnarray}
formed out of the antiferromagnetic (AF) order parameter

\begin{eqnarray}
N_\alpha = \frac{1}{2}\sum_p c_{p+Q}^{\dagger}\sigma^\alpha c_{p} 
\end{eqnarray}
 and the real and imaginary components of the $d$-wave superconducting 
(dSC) order parameter 
\begin{eqnarray}
 \Delta = \sum_p g_p c_{-p\downarrow} c_{p\uparrow}. 
\end{eqnarray}
 Here, 
\begin{eqnarray}
 g_p = \frac{1}{\sqrt{2}} ( cosp_x - cos p_y )
\label{gp}
\end{eqnarray}
is the $d$-wave form factor. 
 The $\pi$ operator rotates $N_\alpha$ and $\Delta$ into each other
\begin{eqnarray}
\left[ \pi_{\alpha}, N_{\beta} \right] = 
 i \Delta~\delta_{\alpha \beta}
\label{commute}
\end{eqnarray}
therefore within the $SO(5)$ theory, AF and dSC are unified into a common
object, called superspin, which can be pictured as a unit vector on an
$SO(5)$ sphere, see Fig.\ref{pi-sphere.eps}. 
 A direct first-order transition between these two phases can
be induced by a chemical potential $\mu$, and the superspin flops 
from the AF direction into the dSC direction. 
 However, inside the dSC phase, there are 4 collective modes, which can
be viewed as Goldstone modes of the spontaneous $SO(5)$ symmetry breaking. 
 The usual SC phase mode corresponds to the rotation inside the dSC plane, 
while there are three extra $\pi$ modes, corresponding to rotations towards 
the AF directions, see Fig.\ref{pi-sphere.eps}. 
 Because $\mu$ breaks the $SO(5)$ symmetry explicitly and constrains the 
superspin to lie at the equator, the $\pi$ fluctuations are massive. 
 From this general consideration, we expect its mass, or the resonance 
energy, to decrease with decreasing doping. 

 The $SO(5)$ theory therefore provides a simple picture of the $\pi$
modes as collective rotations in the $SO(5)$ sphere. 
 Without going into the microscopic details, this picture immediately 
provides us with some useful quantitative information. 
 Inside the dSC phase, the right hand side of the operator equation 
(\ref{commute}) can be replaced by the expectation value of the dSC 
order parameter, giving
\begin{eqnarray}
\left[ \pi_{\alpha}, N_{\beta} \right] = i
\langle \Delta \rangle ~\delta_{\alpha \beta}
\label{exp-commute}
\end{eqnarray}
 This equation can be simply interpreted as the commutation relation between 
a set of canonically conjugate variables, just like $p$ and $q$ in 
elementary quantum mechanics. 
 Therefore, we see that a new set of collective quantum degrees of freedom
emerges in the broken symmetry state characterized by a dSC order $\Delta$. 
 This simple consideration explains why the $\pi$ resonance mode is only 
observed below $T_c$. 
 However, in the regime where a fluctuating $\Delta$ order parameter exists, 
the $\pi$ resonance can also appear as a broad feature. 

 In the dSC state, the $\pi$ mode couples directly to the spin operator 
$\vec S_Q = \vec N$. 
 What is the dynamics associated with the $\pi$ operator?  
 This is a model dependent microscopic question. 
 If one is dealing with an $SO(5)$ symmetric microscopic 
model\cite{silvio,henley,burgess}, 
the dynamics of the $\pi$ operator is determined by the equation
\begin{eqnarray}
\left[ {\cal H}, \pi^{\dagger}_{\alpha} \right] = \omega_0
\pi^{\dagger}_{\alpha}
\label{H-pi-com}
\end{eqnarray}
where $\omega_0=-2\mu$. 
 Therefore, $SO(5)$ symmetric models predict a sharp $\pi$ resonance 
whose energy scales with doping. 
 The dynamics associated with the coupled $\pi$ and spin operator 
in the $t$-$J$ model is the central question studied in our current paper. 
 However, even without detailed microscopic calculations, we 
can give general arguments to estimate the contribution from the
$\pi$ operator to the spin correlation function.
 Equation (\ref{exp-commute}) leads to an important 
sum rule for the mixed correlation function between the spin and the $\pi$
operators. 
 Defining the mixed correlation function as
\begin{eqnarray}
m_{\alpha\beta}(\omega) &=& - \langle 0 | \pi_{\alpha} 
     \frac{1}{\omega - {\cal H} + E_0 + i 0} N_{\beta} 
   - N_{\beta} \frac{1}{\omega - {\cal H} + E_0 + i 0} 
   \pi_{\alpha} | 0 \rangle 
\end{eqnarray} 
 and making use of eq.(\ref{exp-commute}), we have
\begin{eqnarray} 
\int \frac{ d \omega}{2 \pi} 
m_{\alpha\beta}(\omega) &=&
-~\delta_{\alpha \beta} 
\langle 0 | \Delta  | 0 
\rangle 
\end{eqnarray} 
 In addition, we also have another sum rule for the $\pi$ correlation 
function, which follows from the commutation relation
\begin{eqnarray}
\left[ \pi_{\alpha}, \pi_{\beta}^{\dagger} \right] = (1 - n )
~\delta_{\alpha  \beta} 
\end{eqnarray}
where $n$ is a filling factor (half filling corresponds to $n=1$). 
 From these two sum rules, we can put a lower bound on the $\pi$ 
contribution to the spin excitation spectrum as
\begin{eqnarray}
  I_\pi \equiv \frac{1}{\pi} \int_{\pi \ {\rm peak}} d\omega  
     {\rm Im}\chi^{+-}(Q, \omega ) 
    \gsim \frac{2|\Delta|^2 }{1-n}
\label{lower-b1}
\end{eqnarray}
as shown in the Appendix.
 Therefore, one would expect the $\pi$ contribution to the dynamic 
spin correlation function to scale as the square of the dSC order 
parameter and inversely with the doping $x=1-n$. 
 Both of these conclusions are consistent with the experimental finding 
in the optimally doped $YBCO$ that the neutron resonance mode disappears above 
$T_c$ as a sharp excitation, and with the doping dependence of its intensity.
 We can use typical values of $\Delta_{(\pi,0)} = 40 meV$ 
(see Appendix for converting the order parameter to the
quasiparticle energy gap), $J=120 meV$ and doping $x=15\%$ to estimate
the lower bound for $I_\pi$ as 0.32. 
 This is close to and consistent with the 
experimentally measured value, $0.51 \pm 0.1$\cite{fong3}. 
 In a realistic model, the $\pi$ operator is not an exact eigenoperator
of the Hamiltonian, and eq.(\ref{H-pi-com}) is only approximately 
fulfilled with other contributions to the energy $\omega_0$. 
 However, as long as the $\pi$ operator remains as an approximate 
eigenoperator, it will make a sharp contribution to the spin correlation 
function, and the energy of the mode will have a leading contribution 
of $-2\mu$.

\section{ Self-Consistent Formalism }
\label{CF}
 In this paper, we shall study the $\pi$ mode of the $t$-$J$ model 
with nearest-neighbor (n.n.) hopping.
 Before presenting the details of the formalism, we would like to answer 
some general questions regarding the use of this model and the approximations.

The first question concerns the effect of the n.n.~Coulomb 
interaction $V\sum_{i,j}n_i n_j$ \cite{baskaran}. 
 Even if we did not include a bare $V$ term, the reduction from the 
on-site Hubbard model to the 
$t$-$J$ model would generate such a term with $V=-J/4$. 
 Actually, at this particular value of $V$, the interaction between the
quasi-particles making up a spin triplet is zero. 
 One might be concerned that without the multiple scattering in the 
triplet channel, there would not be any $\pi$ resonance.  
 However, this is not the case.  
 Even in the absence of a triplet interaction, there is a sharp $\pi$ mode 
given by 
\begin{eqnarray}
\int dt e^{i \omega t} \theta(t) \langle 0 | \pi_{\alpha}(t)
\pi^{\dagger}_{\beta}(0) | 0 
\rangle  = \frac{1 - n }{\omega + 2 \mu + i 0 }~\delta_{\alpha \beta} .
\end{eqnarray}
 This occurs because the p-p continuum collapses to a point at total 
momentum $q=Q$.  
 Interaction in the triplet channel simply shifts the resonance
energy from $-2 \mu$.  
 In this paper, we shall only calculate the $\pi$ and spin correlation 
function with $V=0$.  
 The effect of $V$ is two-fold, it changes {\it both} the interaction 
in the $\pi$ triplet channel and the energy to destroy a $d$-wave pair 
in the ground state, thereby changing the chemical potential $\mu$.
 Since the $V$ interaction does not distinguish between the singlet Cooper 
pair and the triplet $\pi$ pair, these two contributions will essentially 
cancel each other. 
 This cancelation is indeed observed rather accurately in the numerical
calculations in both the Hubbard and the $t$-$J$ model\cite{meixner,eder}. 
 Because the $J$ interaction is different in the singlet and triplet
channels, this cancelation does not occur.
 Therefore, in this paper, we shall only study the effect of the $J$ 
interaction. 

 The second question concerns the effect of the next nearest neighbor
hopping term $t'$\cite{meixner,eder,baskaran,review}. 
 In the presence of this term, the p-p continuum no longer 
collapses at total momentum $Q$, and it is not clear if
the $\pi$ mode can remain sharp in the presence of $t'$. 
 This question depends on the bandwidth around the $(\pi,0)$ and $(0,\pi)$ 
points in reciprocal space. 
 While the bare bandwidth might be large, it is known from both 
photoemission and numerical experiments that many-body corrections 
reduce the bandwidth at these points significantly.
 Assuming the reduced band structure, the $\pi$ mode remains sharp
in the normal state. 
 Direct numerical calculations on the $\pi$ resonance also show that the 
$\pi$ mode remains sharp for a wide range of $t'$ \cite{meixner,eder}. 
 Because the many-body reduction of the bandwidth is hard to obtain from 
direct perturbation theory, we shall not address the $t'$ issue in this paper.

In this work, we shall mainly discuss the two-dimensional case where
the $\pi$ operator carries momentum $(\pi,\pi)$. 
 Generalizations to bilayer system is straightforward. 
 In this case, the $\pi$ operator rotates the 3D AF state into the 3D dSC 
state, and carries momentum $(\pi,\pi,\pi)$, i.e. it is odd under bilayer 
interchange. 
 If the 3D $\pi$ operator is an approximate eigenoperator of the 
inter-layer Hamiltonian, analysis presented in this paper will carry 
through in the bilayer case as well.

Finally we would like to address the issue of the large Hubbard $U$
repulsion or the no double occupancy constraint in the $t$-$J$ 
model\cite{greiter,reply}.
 In this paper, we shall only treat the Hubbard $U$ within the Hartree 
approximation. 
 In this case, its effect can be captured by a renormalization of the 
chemical potential\cite{reply} and the hopping $t$. 
 Alternatively, we can treat the $t$-$J$ model within the slave boson mean 
field theory. 
 Here one replaces the electron operator $c_{i\sigma}$ by a product of 
$b_i f_{i\sigma}$.
 Within the dSC state, the holons $b_i$ are condensed and can be replaced 
by its c-number expectation value. 
 The resulting Hamiltonian for the spinons $f_{i\sigma}$ is just a 
$t$-$J$ model with renormalized parameters,
where the constraint is only treated on the average, again by adjusting 
the chemical potential and renormalizing the hopping
parameter\cite{tsvelik}. 
 These two formalism therefore lead to the same perturbation series 
in the interaction $J$.

We now review the self-consistent formalism for computing the spin 
correlation function in the SC state. 
 This self-consistent approach has been pioneered by Anderson\cite{anderson} 
and Rickayzen\cite{rickayzen} in treating the problem of the response of a 
superconductor to an electromagnetic field and later used by Bardasis and 
Schrieffer to study collective excitations in a superconductor\cite{bardasis}. 
 The basic idea of this method is the same as that of any linear response
calculation. We perturb a system
by a small external field and then compute the corresponding induced
response. 
 It is however important to remember that when the system has
SC order, any fluctuation in the p-h channel
immediately mixes with fluctuations in the p-p, and
hole-hole channels. 
 This mixing is responsible for restoration  of the
transversality of the electromagnetic response of a
superconductor\cite{rickayzen} and preserving the Ward identities
\cite{eta}. 
 Microscopically it corresponds to taking into account the response
of the superconductor due to the backflow of the condensate as well as 
the creation of the  quasi-particle excitations.
 We have applied this formalism to the $\eta$-resonance in the
negative-$U$ Hubbard model\cite{eta} ( see also \cite{kostyrko} ), and
shown that it constitutes a 
conserving approximation, which gives excellent agreement with the exact
theorems on the $\eta$-resonance of the $U<0$ Hubbard
model\cite{yang,zhang}. 
 Similar formalism has been used recently by
Kohno, Normand and Fukuyama\cite{kohno}, Salkola and
Schrieffer\cite{salkola}, and Brinckmann and Lee\cite{brinkmann} to
study collective excitations. 

 In this paper we emphasize that the origin of the neutron resonance
peak is coupling to the p-p channel below $T_c$. 
 The SCLR formalism is a complete framework which takes this effect 
into account, and has been shown to agree with exact theorems where they 
are available\cite{eta}. 
 However, the naive RPA formula
$\chi_{RPA}=\chi_{BCS}/(1+V_Q \chi_{BCS})$ also contains partial
information about mixing into the non-interacting p-p channel
due to the anomalous $F^{\dagger} F$ term in $\chi_{BCS}$. 
 Therefore
the peak observed at $-2 \mu$ in the RPA treatment may also have its
origin due to p-p mixing. 
 This argument is further strengthened by the findings in our present work 
that the RPA peak at $-2 \mu$ moves to the energy of the interacting triplet 
pair within SCLR formalism.

 We start by considering the $t$-$J$ model in the presence of a
magnetic field $ Re~ h_{q \omega} e^{ i ( q x - \omega t )} $ ( only
the Zeeman effect of the applied field is of
interest to us )
\begin{eqnarray}
{\cal H} = - t \sum_{<ij>\sigma} ( c_{i \sigma}^{\dagger} c_{j \sigma} 
  + {\rm h.c.} )
+ J \sum_{<ij>} {\bf S}_i{\bf S}_j - S_{q}^{-} h_{q\omega} e^{-i
  \omega t} - \mu \sum_{i \sigma} c_{i \sigma}^{\dagger} c_{i \sigma}
\label{t-J}
\end{eqnarray}
where 
$S_{q}^{-} = \sum_{p}
 c_{p \downarrow}^{\dagger} c_{q + p \uparrow }$
is the spin operator. 
 The system responds to the applied field
$h_{q\omega}$ in the spin channel as well as in the $\pi$
channels in such a way that the operators 
\begin{eqnarray}
S_q &=& \sum_p c_{q+p \uparrow}^{\dagger} c_{p
  \downarrow} 
\label{Sop}\\
\pi_q^{+} &=& \sum_p g_p  c_{q+p \uparrow}^{\dagger}
  c_{ - p \uparrow}^{\dagger} 
\label{pi+}\\
\pi_q^{-} &=& \sum_p g_p  c_{ - q - p \downarrow}
  c_{ p \downarrow} \label{pi-}   
\end{eqnarray}
 get non-vanishing time-dependent expectation values. 
 Their Fourier transform will be denoted as 
$S_{q\omega}= \int dt e^{i \omega t} \langle S_q(t) \rangle$ and 
$\pi^{\pm}_{q\omega}= \int dt e^{i \omega t} \langle \pi^{\pm}_q(t) \rangle$.
 The weight function, 
$g_p = ( cos p_x - cos p_y )/\sqrt{2}$ (as defined earlier in eq.(\ref{gp})), 
of the $\pi_q^{\pm}$ operators  arises from the assumed $d$-wave symmetry of 
the order parameter\footnote{In principle,
 the symmetry of the interaction leads to responses in the p-h channel 
 that differ from (\ref{Sop}) by the possible symmetry factors
 $ S_q^{\alpha} = \sum_p \alpha_p 
 c_{ p + q\uparrow}^{\dagger} c_{p \downarrow} 
 \label{Salpha} $
 where $\alpha_p$ may be  $( sin p_x \pm sin p_y ) $ or
 $( cos p_x \pm cos p_y ) $. 
  Such $S_{q\omega}^{\alpha}$ fields provide intermediate states that the spin 
 fluctuations can be scattered into. 
  In general this can modify the amplitude of the induced 
 $S_{q\omega}$ field. 
  However it turns out that for the $q$'s of interest near $Q=(\pi,\pi)$, 
 the effect of such $S_{q\omega}^{\alpha}$ fields is negligible.}.

The Hamiltonian (\ref{t-J}) is then linearized by factoring out 
the dSC gap\footnote{$\Delta_0$ is the energy gap, which is related to 
 the order parameter $\Delta$ defined earlier by the relation, 
 $\Delta_0 = V_{BCS} \Delta$ with $V_{BCS} = \frac{3J}{2}$.}  
 $\Delta_p = \Delta_0~g_p$ and the quantities 
 $\pi_{q\omega}^{\pm}$ and $S_{q\omega}$: 
\begin{eqnarray}
{\cal H} &=& \sum_{p \sigma} \epsilon_p c_{p \sigma}^{\dagger} c_{p
  \sigma} + \sum_{p} \Delta_p c_{p \uparrow}^{\dagger} c_{-p
  \downarrow}^{\dagger} + \sum_{p} \Delta_p^{\ast} c_{- p
  \downarrow} c_{ p\uparrow} \nonumber\\
&+&\frac{J}{4} \pi_{q\omega}^{+} e^{-i \omega t} \sum_p g_{p} c_{- p \uparrow}
  c_{q + p \uparrow}  + \frac{J}{4} \pi_{q\omega}^{-} e^{-i \omega t}
  \sum_p g_{p} 
  c_{p \downarrow}^{\dagger}  c_{-q - p
  \downarrow}^{\dagger} \nonumber\\
&+& ( V_q S_{q\omega} - h_{q\omega} ) e^{-i \omega t} \sum_p  c_{p
  \downarrow}^{\dagger} 
  c_{q + p \uparrow } 
  \label{linear}
\end{eqnarray}
where $V_q = J ( \cos q_x + \cos q_y )$ and 
$\epsilon_p = -2 t (\cos p_x + \cos p_y ) - \mu$.
 We proceed by taking the first two terms in eq.(\ref{linear}) as the 
unperturbed Hamiltonian ${\cal H}_0$ and use Kubo formulas to 
treat the last three terms as the perturbation ${\cal H}_1$, 
\begin{eqnarray}
\langle\hat{f}(t)\rangle  = -i \int_{-\infty}^{t} dt' \langle \left[
    \hat{f}(t), {\cal 
    H}_1 (t')\right] \rangle_{{\cal H}_0} . 
\label{Kubo}
\end{eqnarray}
 This procedure is described in detail in our earlier paper on the $\eta$ 
excitation of the negative-$U$ Hubbard model\cite{eta}. 
  
It is convenient to introduce the amplitude and phase oscillations as  
$b^{+}_{q \omega}= \pi_{q\omega}^{+} + \pi_{q \omega}^{-}$ and 
$b^{-}_{q \omega}= \pi_{q\omega}^{+} - \pi_{q \omega}^{-}$. 
 After some simple calculations, we arrive at the coupled equations for 
$b^{+}_{q \omega}$, $b^{-}_{q \omega}$ and $S_{q \omega}$: 
\begin{eqnarray}
b^{+}_{q \omega} &=& \frac{J}{4} t_{++} b^{+}_{q \omega} + \frac{J}{4}
t_{+-} b^{-}_{q \omega} - 2 V_{q} m_{+} 
( S_{q \omega} - h_{q \omega}/V_{q}) \nonumber\\ 
b^{-}_{q \omega}&=& \frac{J}{4} t_{+-} b^{+}_{q \omega} +\frac{J}{4}
t_{--} b^{-}_{q \omega} -2 V_q m_{-} 
( S_{q \omega} - h_{q \omega}/V_q) \\
S_{q \omega}&=& - \frac{J}{4} m_{+} b^{+}_{q \omega} -\frac{J}{4}
m_{-} b^{-}_{q \omega} -V_q \chi_{0} 
( S_{q \omega} - h_{q \omega}/V_q) 
\label{set}
\end{eqnarray}
 where 
\begin{eqnarray}
t_{++}=& i& \sum_p g_p^2 \int \frac{ d \nu}{2 \pi} 
 \{G_{-p }(\nu - \omega)  G_{ p + q}( - \nu )
 + G_{-p }(\nu )  G_{ p +q}( \omega - \nu ) \nonumber\\
&+& 2 F_{-p } (\nu - \omega)  F_{p+q}(-\nu ) \} \nonumber\\
=&2 &\sum_p g_p^2 ( u_{-p} u_{ p+q} +
 v_{-p} v_{ p+q})^2 
\frac{  \nu_{pq}}{\omega^2 - \nu_{pq}^2}  \nonumber\\
t_{+-}=& i& \sum_p g_p^2 \int \frac{ d \nu}{2 \pi} 
 \{G_{-p }(\nu - \omega)  G_{ p + q}( - \nu )
 - G_{-p }(\nu )  G_{ p + q}( \omega - \nu ) \} \nonumber\\
=&2& \sum_p g_p^2 ( v_{-p }^2v_{ p+q}^2 -  u_{-p}^2 u_{ p + q}^2 ) 
     \frac{  \omega}{\omega^2 - \nu_{pq}^2} \nonumber\\ 
t_{--}=& i& \sum_p g_p^2 \int \frac{ d \nu}{2 \pi} 
\{ G_{-p }(\nu - \omega)  G_{ p + q}( - \nu )
 + G_{-p }(\nu )  G_{ p + q}( \omega - \nu ) \nonumber\\
 &-& 2 F_{-p } (\nu - \omega)  F_{ p+q}(-\nu ) \} \nonumber\\
=&2 &\sum_p g_p^2 ( u_{-p} u_{p+q} - v_{-p} v_{p+q})^2 
 \frac{  \nu_{pq}}{\omega^2 - \nu_{pq}^2}  \nonumber\\
m_{+}=& i& \sum_p g_p \int \frac{ d \nu}{2 \pi} 
 \{ F_{p + q} ( \nu )  G_{ -p }( \nu+\omega) 
 - F_{- p } ( -\nu - \omega )  G_{ p + q}( \nu ) \} \nonumber\\
= &2& \sum_p g_p u_{ p +q} v_{ p +q} (u_{-p}^2 -  v_{ -p }^2 ) 
 \frac{\nu_{pq}}{\omega^2 - \nu_{pq}^2}  \nonumber\\ 
m_{-}= &-& i \sum_p g_p \int \frac{ d \nu}{2 \pi} 
\{ F_{p +q} ( \nu  )  G_{ -p }( \nu+\omega ) 
 + F_{-p } (- \nu - \omega )  G_{ p + q}( \nu ) \} \nonumber\\
=  &-&2 \sum_p g_p u_{p + q } v_{p + q} 
 \frac{\omega}{\omega^2 - \nu_{pq}^2}  \nonumber\\
\chi_0 =&i& \sum_p  \int \frac{ d \nu}{2 \pi} 
 \left\{ G_{-p}(\nu ) G_{ p+q}( \nu + \omega ) 
       + F_{-p}(-\nu) F_{ p+q}( \nu -\omega ) \right\} \nonumber\\
=&- &\sum_p ( u_{p+q} v_{ -p} - u_{-p} v_{ p+q})^2 
\frac{\nu_{pq}}{\omega^2 - \nu_{pq}^2} 
\label{factors} 
\end{eqnarray}
 In the equations above, 
$E_p = \sqrt{ \epsilon_p^2+ \Delta_p^2}$,~
$\nu_{pq} = E_{p+q}+E_{-p}$,~ 
$u_p v_p = \Delta_p/ 2 E_p $,~ 
$u_p^2 = \frac{1}{2} (1 + \frac{\epsilon_p}{E_p})$, and 
$v_p^2 = \frac{1}{2} (1 - \frac{\epsilon_p}{E_p})$. 
 The Green's functions have been defined as 
\begin{eqnarray}
  G_p(\omega) &=& \int dt e^{i 
  \omega t} ( - i) \langle T c_{p\sigma}(t) c_{p\sigma}^{\dagger}(0) \rangle 
\nonumber\\
  F_p(\omega) &=& \int dt e^{i \omega t} ( - i) \langle T c_{p\uparrow}(t)
  c_{-p\downarrow}(0) \rangle \nonumber\\  
 F_p^{\dagger}(\omega) &=& \int dt e^{i
  \omega t} ( - i) \langle T c^{\dagger}_{-p\downarrow}(t)  
  c^{\dagger}_{p\uparrow}(0) \rangle 
\end{eqnarray}
 In eq.(\ref{factors}), $\omega$ should be taken to have an infinitesimal 
imaginary part, $\Gamma = 0^+$, coming from causality.

\begin{figure*}[h]
\centerline{\epsfxsize=12cm
\epsfbox{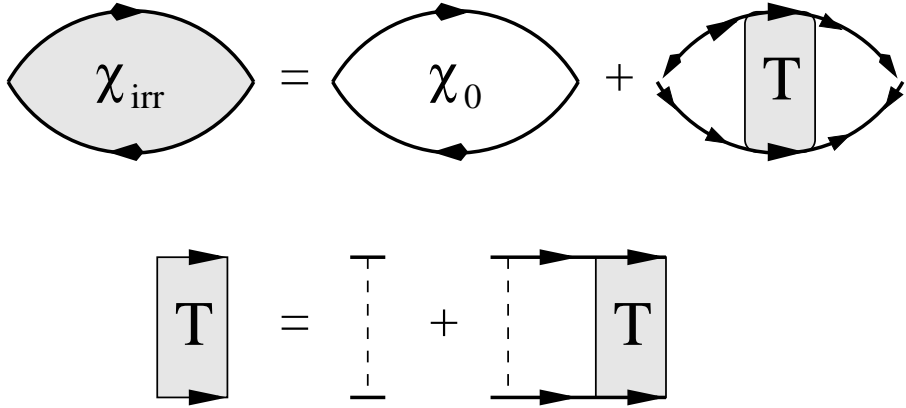}
}
\caption{ Modification of $\chi_{irr}$ due to ladder diagrams }
\label{diag.eps}
\end{figure*}

Solution of eqs.(\ref{set}) gives the dynamical spin susceptibility
in this Self-Consistent Linear Response Theory ($SCLR$). 
\begin{eqnarray}
\chi_{SCLR}(q,\omega) 
  = i \int dt e^{i \omega t} \theta(t) 
      \langle \left[ S_{q}^{+}(t), S_{-q}^{-}(0) \right] \rangle 
  = \frac{S_{q\omega} }{ h_{q \omega} } . 
\label{chi-def}
\end{eqnarray}
 It can be written in the form
\begin{eqnarray}  
 \chi_{SCLR}(q,\omega) 
& = & \frac{ \chi_{irr}}{1 + V_q \chi_{irr} }
\label{mo-rpa}
 \\
 \chi_{irr}&=&\chi_0 + \Delta \chi 
\label{chi_irr} \\
\Delta \chi & = & - \frac{J}{2}~ \frac{ m_{+}^2 + m_{-}^2 -\frac{J}{4}
  m_{+}^2 t_{--} - \frac{J}{4}m_{-}^2 t_{++} + \frac{J}{2} m_{-} m_{+}
  t_{+-} }{ 1 - \frac{J}{4} t_{++} - 
  \frac{J}{4} t_{--}  +\frac{J^2}{16} t_{++}t_{--} - \frac{J^2}{16}
  t_{+-}^2}
\label{MRPA}
\end{eqnarray} 
and may be understood as a modified Random Phase Approximation (RPA) where
the bare bubble $\chi_0$ has been modified by including the ladder
diagrams. 
 Fig.\ref{diag.eps} gives the diagrammatic interpretation
of formulas (\ref{chi_irr}), (\ref{MRPA}).

The procedure for finding the p-p correlation function 
\begin{eqnarray}
P(q,t) = -i \theta(t) \langle \left[ \pi^{+}_q(t), \pi^{-}_q (0)
\right] \rangle
\end{eqnarray} 
is very similar to the one shown above for the spin channel. 
 We only need to add an external field in the $\pi_Q^{\dagger}$ channel 
and compute the response in the same channel. 
 Skipping the laborious but straightforward calculations we present the 
final expression for its Fourier transform 
\begin{eqnarray}
 P(q, \omega)= \frac{t_{++}+t_{--}+\frac{J}{2}t_{+-}^2
 -\frac{J}{2}t_{++}t_{--}-2 t_{+-} -\frac{J}{4} 
\frac{ m_{+}^2 ( 1- \frac{J}{2} t_{--} ) + m_{-}^2 ( 1 - \frac{J}{2}
t_{++} ) -2 m_{+} m_{-} (1 - \frac{J}{2}t_{+-}) } { 1 + V_q \chi_0} }
{1- \frac{J}{4} t_{++} - \frac{J}{4} t_{--} -\frac{J^2}{16} t_{+-}^2
 +\frac{J^2}{16} t_{++} t_{--} + J^2 
 \frac{m_{+}^2(1-\frac{J}{4}t_{--}) + m_{-}( 1-\frac{J}{4}t_{++})
+\frac{J}{2}m_{+} m_{-} t_{+-} }{ 1 + V_q \chi_0 } }
\end{eqnarray}

In the normal state this reduces to a simple T-matrix expression that
was studied in \cite{PRL}. 
 There it was shown that, in the normal state, the p-p spectrum at $Q$ 
is dominated by the collective $\pi$-mode resonance that appears due to 
the collapse of the p-p continuum 
($\epsilon_{p + Q}+\epsilon_{-p} = -2 \mu$ ) and the repulsive
interaction of two particles in a triplet state sitting on n.n.~sites. 
 We suggested that this collective mode may contribute to the
spin-fluctuation spectrum when the system becomes superconducting. 
 However such argument raises an immediate concern that 
superconductivity could in principle lead to another effect $-$ a 
significant broadening of the $\pi$ resonance, due to possible scattering 
into the p-h excitations. 
 The goal of the next part is to show that this does not happen. 
 The $\pi$ resonance survives as a collective mode and affects strongly 
the dynamic spin-spin correlation function in the SC state. 
 The important point here is that unlike $\chi_0$, $\Delta \chi$ in
$\chi_{irr}$ contains information about the $\pi$ resonance. 
 As we shall see in the next section, $Im~\chi_{irr}$ nearly vanishes at the
$\pi$ resonance energy, where $Re~\chi_{irr}$ is sharply peaked. 
 The combination of these two effects give rise to a sharp $\pi$ resonance.

\section{Numerical Results}
\label{numerix}

It is well known that  the RPA form of the spin correlation function 
overestimates the antiferromagnetic instability. 
 Therefore, if we see a peak in the dynamic spin correlation function, 
it is important to check if it is an artifact due to the RPA type of 
overestimate or due to a genuine collective mode.
 Moreover, the size of the dSC gap relative to $T_c$ is
significantly larger than the BCS estimate. 
 The BCS gap equation for
${\cal H}_0$ is
\begin{eqnarray}
 1 = V_{BCS} \sum_p \frac{g_p^2}{2 E_p} tanh (\frac{E_p}{2T}) 
\label{gap-eq}
\end{eqnarray}
with the bare pairing interaction $V_{BCS}= 3J/2$. 
 However, this gives a $2\Delta_{(\pi,0)}/k_B T_c$ ratio of the order of 
4 which is small compared to the typically observed value of 6 to 8. 
 Therefore, in what follows, we take two approaches to these problems. 
 We introduce an effective reduction of the antiferromagnetic vertex 
$V_{\bf Q}=\alpha V^{bare}_{\bf Q}$ with $\alpha < 1$, as a way to model 
vertex corrections, or we take the dSC gap $\Delta_0$ to be 
bigger than its mean field value. 
 Both of these approaches have the effect of removing the RPA type of AF 
instability. 
 We shall see that the $\pi$ resonance is robust against these variations.

\subsection{ $\pi$ resonance and its robustness against vertex corrections }

In this section we take $J=0.6t$ and $\mu= -0.3 t$. 
 We choose the mean-field value of $\Delta_0=0.0094 t$ and the reduction 
of $V_Q$ is set by $\alpha = 0.82$. 
 We assume a finite value $\Gamma=10^{-4}t$ for the imaginary infinitesimal 
in the energy denominator and perform integration by dividing the
Brillouin zone into a $32000\times32000$ lattice.

\begin{figure}[hbt]
\epsfbox{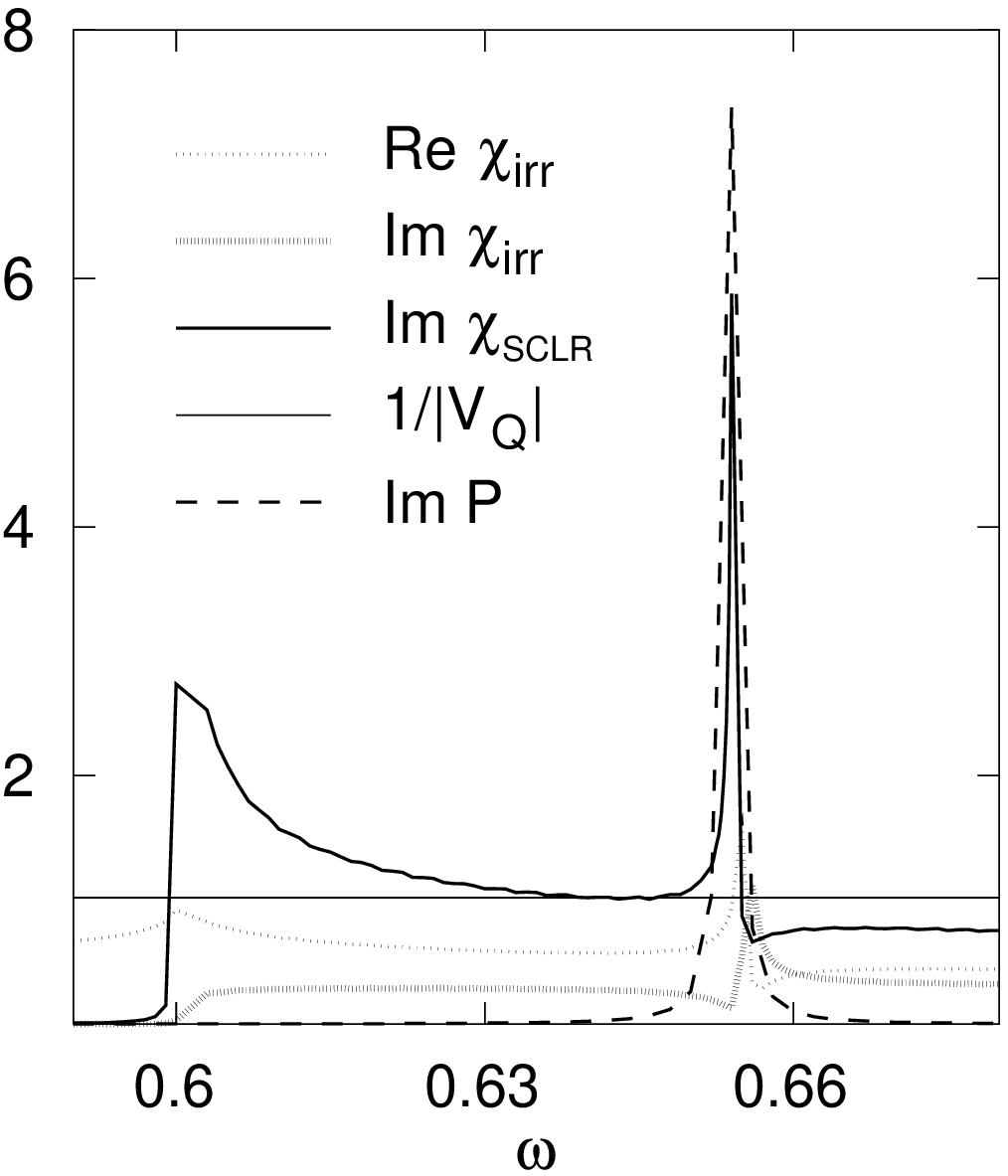}
\hspace{0.5cm}
\caption{ Re $ \chi_{irr}$, Im $ \chi_{irr}$, Im $ \chi_{  SCLR}$, 
  and Im$P$ vs.~$\omega$ } 
\label{Mxcomb.eps}
\end{figure}

This plot shows the ``mechanics'' of the resonance in $\chi_{SCLR}$. 
 In the normal state the p-p channel has a sharp peak at 
$\omega_0 \approx -2 \mu + \frac{J}{2}(1-n) = 0.655t$.
 Notice that there is no visible shift of the energy of this resonance
in the SC state, but only a small broadening.
 This resonance in the p-p channel $P(Q,\omega)$ then leads to a peak in 
$Re \chi_{irr}$. 
 Consequently at a frequency where 
 $Re \chi_{irr} = \left|\frac{1}{V_{Q}}\right|$ the real part of the 
denominator in the SCLR expression (\ref{mo-rpa}) vanishes leading to a
peak in $ Im \chi_{SCLR}$. 
 At these frequencies, the imaginary part of the denominator 
($Im \chi_{irr}$ ) is also small, and the resonance appears to be quite sharp.

\begin{figure}[hbt]
\epsfbox{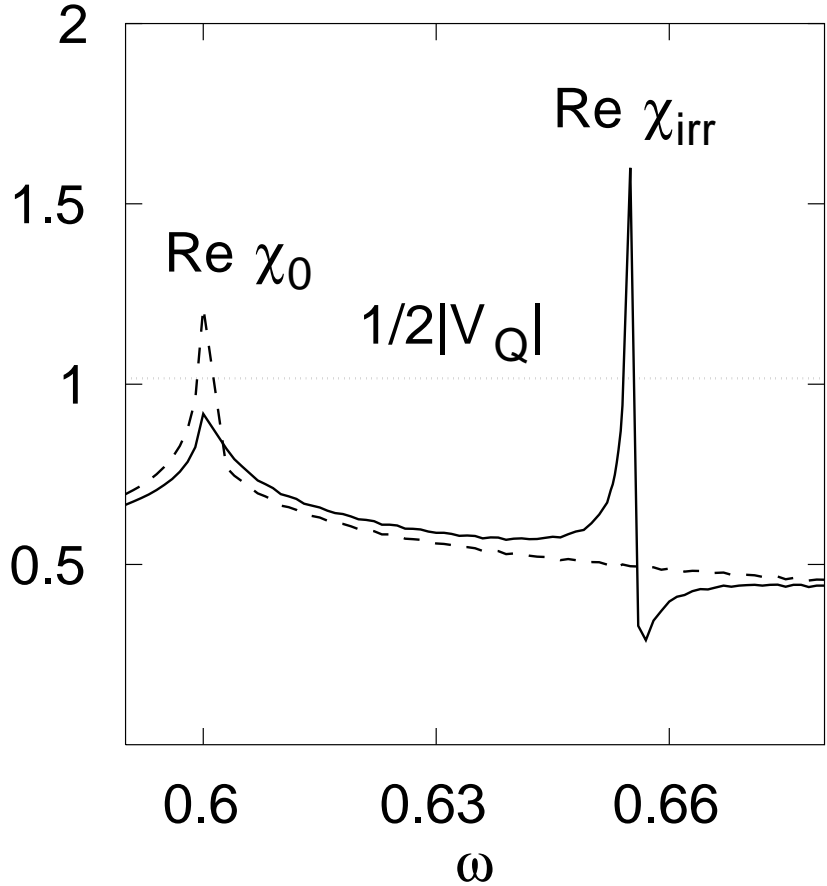}
\caption{ Re$\chi_{irr}$ and Re$\chi_0$ vs.~$\omega$. 
    The dotted line represents the line of $1/2J$. }
\label{Mre.eps}
\end{figure}
\begin{figure}[hbt]
\epsfbox{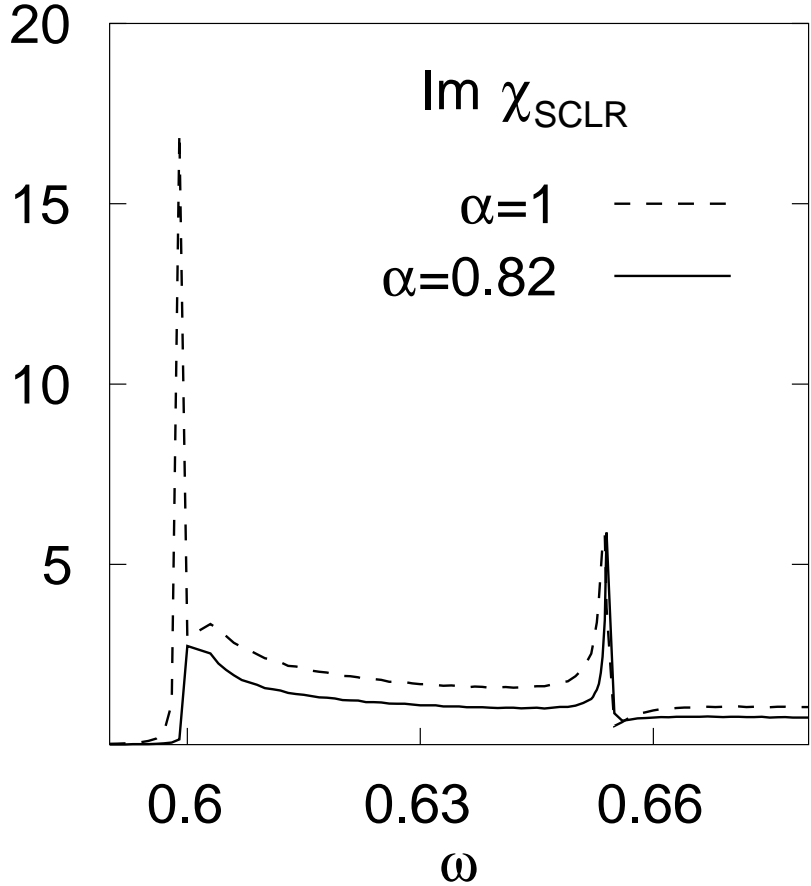}
\vspace{0.1cm}
\caption{ Im $\chi_{SCLR}$ vs.~$\omega$ for two different values of
  vertex correction parameter $\alpha$. }
\label{Malpha.eps}
\end{figure}

 In Fig.\ref{Mre.eps}, we compare the real part of $\chi_{irr}$ with
that of $\chi_0$. 
 As discussed above, we have resonance peaks in $Im \chi_{SCLR}$ 
when $Re \chi_{irr} =\left| \frac{1}{V_Q} \right|$.  
 We can see that taking $\chi_{irr}$ instead of $\chi_0$ considerably
suppresses the divergence around $-2 \mu$ 
(this divergence comes from the dynamic nesting of the Fermi surface. 
 It gives rise to the RPA peak, the only resonance one gets from a naive 
 RPA calculation.)
and leads to the development of a peak at the energy of the $\pi$-excitation. 
 It is easily noticeable that if we do not take into account reduction of 
$V_Q$, but exploit the bare value of $V^{bare}_Q=-2J$, 
then Re$\chi_{irr}$ will cross it at two points 
($\omega \sim -2\mu$ and $\omega_0$), giving rise to both 
$-$ RPA and $\pi$ peaks (see Fig.\ref{Malpha.eps}). 
 However, the divergence of Re$\chi_{irr}$ around $\omega_0$ is much
stronger, making the $\pi$ peak more robust against variations in $V_Q$.

\begin{figure}[hbt]
\epsfbox{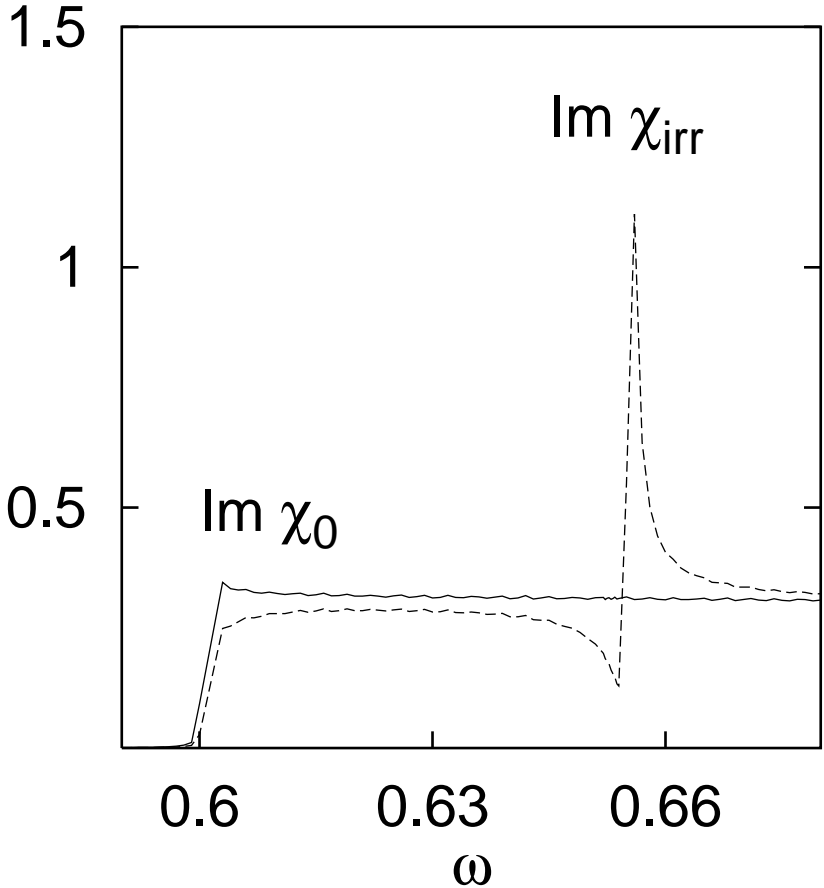}
\caption{ Im $ \chi_{irr}$ and Im $ \chi_0$ vs $\omega$ }
\label{Mim.eps}
\end{figure}

In Fig.\ref{Mim.eps}, we show the Imaginary part of $\chi_{irr}$ and
$\chi_0$.  
 Note that a dip develops in Im$\chi_{irr}$ at the energy of the 
$\pi$-excitation. 
 This means that the $\pi$-resonance is much less damped than one 
might have expected. 
 In the normal state, the stability of the $\pi$ resonance is guaranteed
by the absence of the phase space available for decay (p-p continuum
collapses to a point).
 In the dSC state, this argument no longer works. 
 Mixing of the p-h and p-p channels could provide a mechanism for the 
decay of the $\pi$ excitation. 
 However, we see that the system accommodates the $\pi$
excitation by suppressing Im$\chi_{irr}$ at its energy. 
 In Section \ref{analyt}, we shall give an approximate analytical 
derivation of this important feature.

\begin{figure}[hbt]
\epsfbox{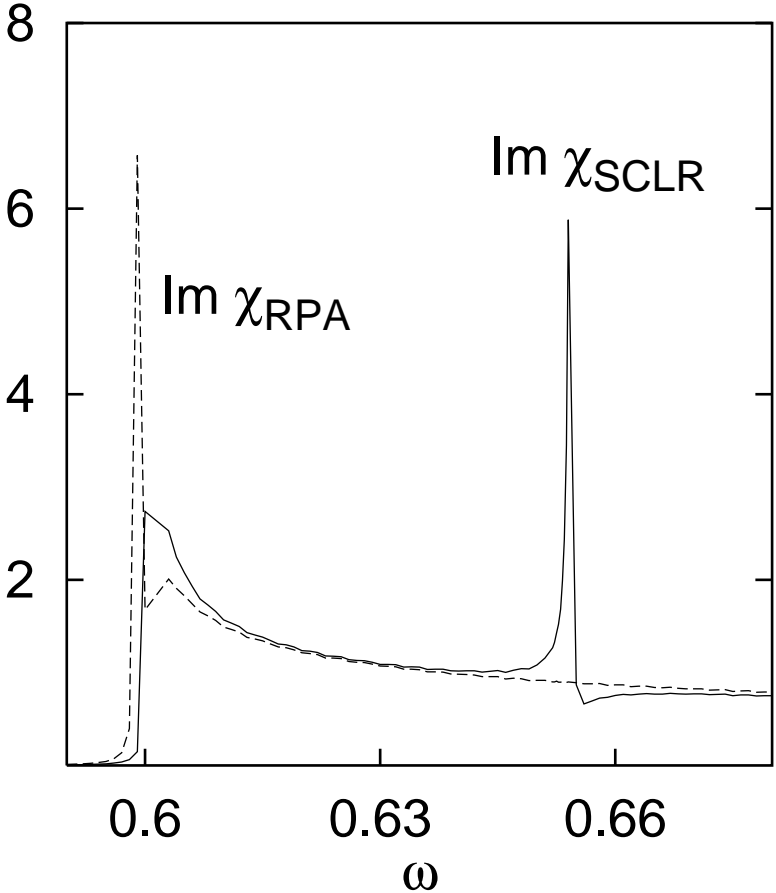}
\caption{ Im $ \chi_{SCLR}$ and Im $\chi_{RPA}$ vs.~$\omega$ }
\label{Mtot.eps}
\end{figure}
 In Fig.\ref{Mtot.eps}, we compare the self-consistent spin-spin
correlation function $\chi_{SCLR}$ with the one obtained from the RPA 
calculation, $\chi_{RPA}$. 
 The latter one has an RPA peak that comes from the dynamic nesting of the 
tight-binding Hamiltonian at momentum $Q$. 
 In Im$\chi_{SCLR}$, this peak disappears almost completely, 
and the the spectral weight is transfered into the $\pi$-excitation.

In Fig.\ref{Malpha.eps}, we show the comparison of different 
choices of $\alpha$ in $V_Q$. 
 Notice the coexistence of the RPA peak with the $\pi$ peak for the 
choice of bare parameter ($\alpha =1$). 
 Reducing $\alpha$ has no effect on the $\pi$ resonance but completely 
destroys the RPA peak.

From the analysis carried out in this subsection, we conclude that the
RPA peak might be the the result of overestimating the AF instability,
while the $\pi$ peak is robust against vertex corrections.

\subsection{ Robustness of the $\pi$ peak against variations of
  the superconducting gap } 

Another way of suppressing AF instability within RPA or SCLR formalism
is to choose a larger dSC energy gap.

\begin{figure}[hbt]
\epsfbox{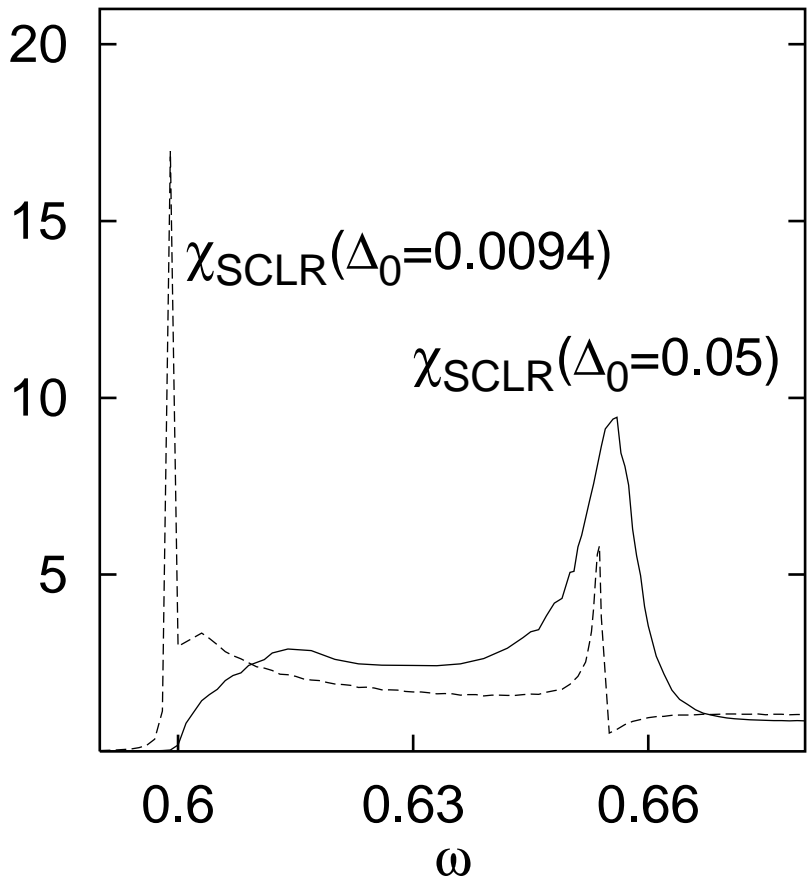}
\caption{ Im$\chi_{SCLR}$ vs.~$\omega$ for different values of $\Delta_0$.}
\label{del.eps}
\end{figure}

In Fig.\ref{del.eps}, we compare the results of SCLR calculations
 for the spin correlation function for two choices of $\Delta_0$. 
 The smaller one, $\Delta_0 = 0.0094t$, corresponds to the self-consistent 
(mean-field) value, and the bigger one was taken as $\Delta_0=0.05 t$. 
 In these calculations, we take $J=0.6t$ and $\mu= -0.3t$ as before, but 
with $V_Q=-2J$, the bare value ($\alpha =1$).

 We observed in the previous subsection that two peaks (RPA and $\pi$) 
coexist with the choice of the mean-field value for $\Delta_0$ and a bare 
value for $V_Q$. 
 Fig.\ref{del.eps} shows that taking a larger dSC gap removes the RPA peak 
and increases the spectral weight of the $\pi$ peak.  
 This has an even stronger effect than we saw in the previous section 
by reducing the AF exchange constant. 
 The latter one, as we found, only removes the RPA peak without affecting 
the $\pi$ resonance. 
 It is also interesting to find that for the larger gap there is an 
increase in the energy of the resonance.

\begin{figure}[hbt]
\epsfbox{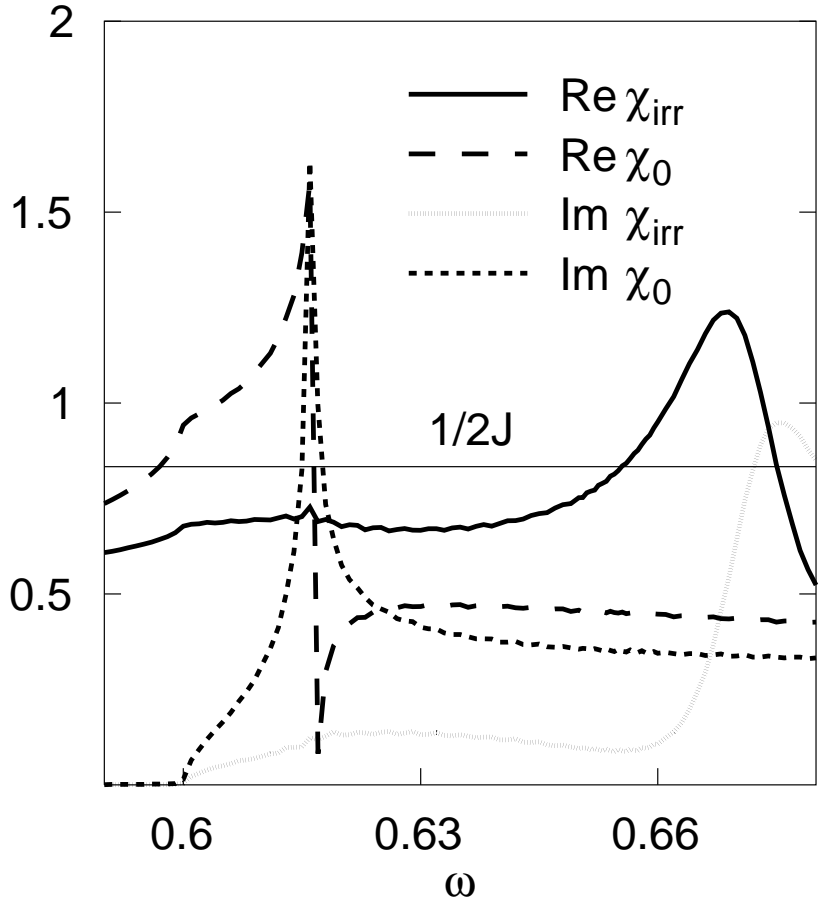}
\caption{  $\chi_{irr}$ and  $ \chi_0$ vs $\omega$ for the case when
  $\Delta_0$ is larger than its BCS value. }
\label{RI.eps}
\end{figure}

A tenacious  effect of the large dSC gap is explained in Fig.\ref{RI.eps}. 
 Here the choice of parameters is the same as in the previous Figure with 
$\Delta_0=0.05t$. 
 By looking at $\chi_{irr}$ in this case of large $\Delta_0$, we find that 
the RPA peak in the real part ($\omega \approx -2 \mu$) has completely 
disappeared.  
 For the mean-field value of $\Delta_0$, there was only a suppression of
this peak. 
 In contrast to that, the only effect of taking a larger $\Delta_0$ on the 
$\pi$ peak in $Re \chi_{irr}$ was to make it broader. 
 This broadening explains the increase in the total weight of the $\pi$
peak in $\chi_{SCLR}$ 
(the slope of $Re \chi_{irr}$ at the crossing point with $1/2J$ determines 
 the total weight of the $\pi$ resonance in $\chi_{SCLR}$. 
 See more on that in Section \ref{analyt}.). 
 Also note an enormous suppression of imaginary part of $ \chi_{irr}$ for 
energies below $\omega_0$ in this case of large $\Delta_0$.

\begin{figure*}[h]
\centerline{\epsfxsize=12cm
\epsfbox{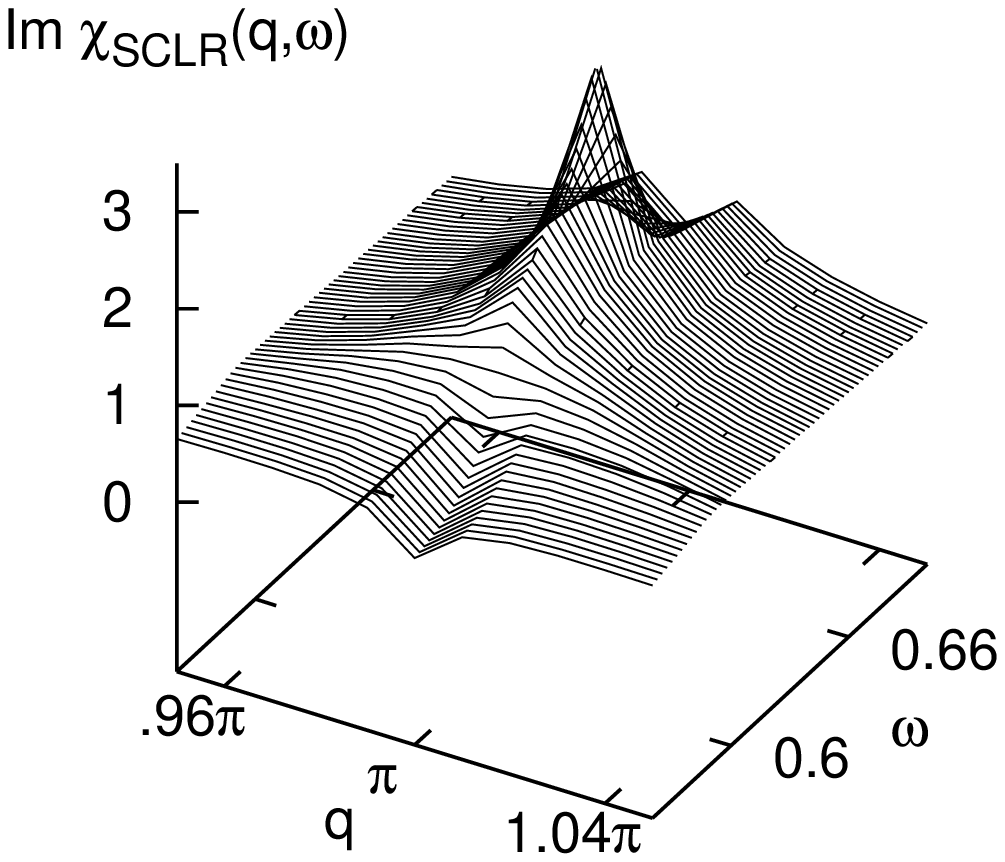}
}
\caption{ Im $ \chi_{  SCLR}({\bf q},\omega)$ for ${\bf q}=(q,q)$ }
\label{qplot.eps}
\end{figure*}

 On Fig.\ref{qplot.eps}, we show the ${\bf q}$ and $\omega$ dependence of 
Im $\chi_{SCLR}({\bf q},\omega)$. 
 This plot has been done for $\Delta_0=0.05t$, $\alpha=1$ and for 
computational reasons we took a larger $\Gamma=10^{-2}t$, which leads to 
considerable smearing. 
 However the general picture of the ${\bf q}$ dependence of 
Im$\chi_{SCLR}({\bf q},\omega)$ may be seen quite clearly. 
 We have an incommensurate structure at low frequencies, which is followed 
by a commensurate $\pi$-peak. 
 Right after the peak, there is a missing spectral weight at the commensurate 
wavevector. 
 Qualitatively this picture is similar to what is seen in inelastic neutron 
scattering in YBCO \cite{dai2}.  
(The incommensurate peaks in that case will be rotated by 45 degrees due to 
 a different band structure. 
 Calculations for the YBCO band structure will be published elsewhere 
 \cite{next}. )

\section{ Analytical derivation of the $\pi$ resonance }
\label{analyt}

In the previous section, we showed numerically that at the energy of the 
$\pi$-excitation,  $\chi_{irr}$ possesses a sharp peak in the real part 
and a dip in the imaginary part.  
 In this section, we study the origin of these properties 
analytically\footnote{We restrict this analysis to  
 the case $q= Q = (\pi,\pi)$.}.   
 When looking at the structure of the expressions in eq.(\ref{set}), 
one encounters very often analogous integrals that only differ by 
a factor $g_p^2$ in the $p$-summation.
 To simplify the following analysis, we make an approximation that this 
factor may be replaced by its average value of 1. 
 A similar assumption has been used in \cite{so5} to obtain approximate 
SO(5) algebra.  
 It is important to realize that one should take the average of $g_p^2$ 
not over the whole Brillouin zone but over a narrow band around the Fermi 
surface, since in most of these expressions the other factors in the 
integrals restrict the important domain of integration to this region.

We introduce 
\begin{eqnarray}
I_1(\omega)& =& \sum_p g_p^2 \frac{ 1 - v_{-p}^2  -
  v_{p+Q}^2 }{\omega^2 - \nu_{pQ}^2}
\nonumber\\
I_2(\omega)& = & \sum_p g_p \frac{ u_{p+Q} v_{p+Q}
  - u_{-p} v_{-p} }{\omega^2 - \nu_{pQ}^2}
\end{eqnarray}
and use some identities for the BCS coherence factors and the 
approximation $g_p^2 \sim 1$ to express all the factors in (\ref{set})
as 

\begin{eqnarray}
t_{++}&=& - 4 \mu I_1(\omega) \nonumber\\
t_{+-}&=& - 2 \omega I_1(\omega) \nonumber\\
t_{--}&=& \frac{1-n}{ \mu} - \frac{\omega^2}{ \mu} I_1(\omega)
\nonumber\\
m_{+}&=&-2 \mu I_2(\omega) \nonumber\\
m_{-}&=& - \omega I_2(\omega) \nonumber\\
\chi_0 &=& \frac{2}{3J} + \frac{ \omega^2 - 4 \mu^2}{2 \Delta_0} I_2(\omega) .
\end{eqnarray}

Substituting these expressions into eq.(\ref{MRPA}) and using the
identity
$- 4 \mu \Delta_0 I_2(\omega) = (1-n) +  (4 \mu^2 - \omega^2 )
I_1(\omega) $ which holds within the approximation described above, we
get 
\begin{eqnarray}
\chi_{irr} = \frac{1}{V_{BCS}} &+& \frac{ I_2(\omega) }{2 \Delta_0}
\frac{ (\omega^2 - \omega_0^2) ( \omega^2 - 4 \mu^2 ) }
{\omega^2 - \omega_0^2 + \Delta_0 J I_2(\omega) (\omega^2 - 2 \mu
\omega_0 ) }
\label{chi_irr_e}
\end{eqnarray}
where $V_{BCS}$ comes from the gap equation
$\sum_p g_p u_{p}
v_{p} = \Delta_0 / V_{BCS}$ (eq.(\ref{gap-eq})), and $\omega_0 = -2 \mu +
  \frac{J}{2} ( 1 - n )$. 
 In the mean field analysis of the $t$-$J$ model, $V_{BCS}=3 J/2$. 
 Since $I_2(\omega)<0$ for $\omega \sim \omega_0$, the denominator of 
the expression (\ref{chi_irr_e}) vanishes when the frequency is larger 
than $\omega_0$ but very close to it (the energy separation is proportional 
to $\Delta_0^2$ ). 
 This explains the peak in Re$\chi_{irr}$, and the factor 
$\omega^2 - \omega_0^2$ in the numerator explains the dip in the imaginary 
part.  

Expression (\ref{chi_irr_e}) allows  us to estimate the integrated spectral
weight of the $\pi$-excitation in the spin-spin correlation function.
 For simplicity, we neglect a small imaginary part of $\chi_{irr}$ near
$\omega_0$. 
 Then a pole in $\chi_{SCLR}$ occurs when 
\begin{eqnarray}
Re \chi_{irr}= - \frac{1}{V_Q} . 
\label{cond}
\end{eqnarray}

Expanding $\chi_{SCLR}$ around this frequency $\omega_{*}$ we find that
\begin{eqnarray}
\chi_{SCLR} = \left. \frac{\chi_{irr}}
 {V_Q \frac{\partial \chi_{irr}}{\partial \omega} } \right|_{\omega_{*}}
\frac{1}{\omega-\omega_{*}+i0} =
 \frac{1} {V_Q^2 \frac{\partial \chi_{irr}}{\partial \omega} |_{\omega_{*}} }
~\frac{-1}{\omega-\omega_{*}+i0 } .
\end{eqnarray}

Earlier we introduced two distinct energies in our system. 
 The BCS coupling $V_{BCS}$ and the AF coupling $V_Q=-2J$.
 If we consider a {\it hypothetical} situation when $V_Q= - V_{BCS}$, 
we can see that the condition (\ref{cond}) is satisfied
exactly at $\omega_0$ and a simple calculation gives
$ \chi_{SCLR}(\omega) = \frac{2\Delta_0^2}{V_{Q}^2 (1-n)}~
\frac{-1}{\omega - \omega_{0}+i0}$.
 If we take here $V_Q$ to be $V_{BCS} = 3J/2$, we have 
for the intensity of the $\pi$ resonance as
\begin{eqnarray}
I_{\pi} = \frac{1}{\pi}\int_{\omega_0 - \nu}^{\omega_0 + \nu'} 
  d\omega \chi''_{SCLR}
= \frac{ 8 \Delta_0^2}{9J^2(1-n)} . 
\label{intens}
\end{eqnarray}
 The right hand side is equal to the expression derived in the 
Appendix as a lower bound.
 In eq.(\ref{intens}), $\nu$ and $\nu'$ characterize the width of the 
$\pi$ resonance around $\omega_0$, and we introduced a factor $1/\pi$ since  
the Lehmann representation of (\ref{chi-def}) is given by 
 \mbox{$Im~ \chi(Q,\omega) = \pi \sum_n \left| \langle n | S_Q^{+} | 0
  \rangle \right|^2 \delta( \omega - \omega_{n0} ) $}.
 This definition of $I_\pi$ is the same as $A(T)$ of 
ref.\cite{fong3}\footnote{ 
 For unit convention, see Appendix B of ref.\cite{fong3}.}. 
 Expression (\ref{intens}) is also what we obtained for the 
intensity of the $\pi$ resonance in \cite{PRL} using a T-matrix 
analysis\footnote{  
 Note a factor of $\sqrt{2}$ difference in the 
 definitions of $\Delta_0$ here and in ref.\cite{PRL}, 
 $\Delta_0({\rm present}) = \sqrt{2} \Delta_0({\rm ref.\cite{PRL}})$.
}.

 If we take realistic values of $J=120~meV$, $\Delta_{(\pi,0)} = 40~meV$ 
(this corresponds to $\Delta_0 = 28.3~meV$) and $1-n=15\%$, and substitute
them into eq.(\ref{intens}), we get
$I_{\pi} = 0.32$. 
 For the $t$-$J$ model, $|V_Q| > V_{BCS}$, and the energy satisfying 
the condition (\ref{cond}) is lower than $\omega_0$ and the slope 
$\frac{\partial \chi_{irr}}{\partial \omega}$ is smaller than the above 
estimate (see Fig.\ref{Mre.eps}).
 This will be partly cancelled with the increse of $V_Q$, and we 
expect eq.(\ref{intens}) gives a semi-quantitative estimate for $I_\pi$.

% This formula shows that the intensity of the $\pi$ peak scales as a
%square of the order parameter $\Delta \sim \Delta_0 / V_{BCS}$ and is
%inversely proportional to doping $x=1-n$. 
% This is in agreement with numerical results of Section \ref{numerix} and 
%is consistent with general model-independent consideration given in the 
%Introduction. 
% If we take realistic values of $J=120~meV$, $\Delta_{(\pi,0)} = 40~meV$ 
%(this corresponds to $\Delta_0 = 28.3~meV$ ) and $1-n=15\%$, and substitute
%them into eq.(\ref{intens}), we obtain 
%$I_{\pi} = 0.185$.

\section{Comparison with the $SO(5)$ Equations of Motion}
\label{EOM}

We now study the Heisenberg equations of motion (EOM) for the $\pi$ and spin 
operators 
\begin{eqnarray}
\pi_{pQ}^+ &=& c_{p+Q \uparrow}^{\dagger} c_{-p\uparrow}^{\dagger}  
\nonumber\\
\pi_{pQ}^- &=& c_{-p-Q \downarrow} c_{p\downarrow} \nonumber\\
S_{pQ}^{+} &=& c_{p+Q \uparrow}^{\dagger} c_{p\downarrow}
\label{opers}
\end{eqnarray}
using (\ref{t-J}) as a Hamiltonian. 
 A closed set of equations may be obtained by taking commutators of 
the operators (\ref{opers}) with the Hamiltonian and then factorizing the 
results in terms of the occupation numbers for the electrons 
$v_p^2 = \langle c_{p\sigma}^{\dagger} c_{p\sigma}\rangle$
and BCS anomalous averages
$u_p v_p = \langle c_{-p\uparrow}c_{p\downarrow} \rangle$. 
 As shown by Anderson and others \cite{anderson}, this procedure recovers 
the Modified Random Phase and T-Matrix Approximations.
\begin{eqnarray}
 \left[ {\cal H}, \pi_{pQ}^+ \right] 
 &=& ( \tilde{\epsilon}_{p+Q} + \tilde{\epsilon}_{-p} ) \pi_{pQ}^+ 
 + \frac{J}{2}~(1 - v_{p}^2 - v_{p+Q}^2)  \sum_k \pi_{kQ}^+ ~\eta(p-k)  
\nonumber\\ 
 &-& \frac{3J}{2} ( S_{pQ}^+ + S_{-p-Q~Q}^+ ) \sum_p u_k v_k \eta(p-k)  
 + 4 J~ u_p v_p \sum_k S_{kQ}^+
\label{pi-d-eom} \\
 \left[ {\cal H}, S_{pQ}^{+} \right] 
 &=& ( \tilde{\epsilon}_{p+Q} - \tilde{\epsilon}_{p} ) S_{pQ}^{+} 
 + 2 J ( v_{p+Q}^2 - v_p^2 ) \sum_k S_{kQ}^{+} \nonumber\\
 &-& \frac{3 J}{2}~( \pi_{pQ}^+ - \pi_{pQ}^- ) \sum_k  u_k v_k~ \eta(p-k) 
  - \frac{J}{2} u_p v_p \sum_k  ( \pi_{kQ}^+ - \pi_{kQ}^- )~ \eta(p-k)
\label{s-eom} \\
 \left[ {\cal H}, \pi_{pQ}^- \right] 
 &=& - ( \tilde{\epsilon}_{p+Q} + \tilde{\epsilon}_{-p} ) \pi_{pQ}^-
  - \frac{J}{2}~(1 - v_{p}^2 - v_{p+Q}^2)  \sum_k \pi_{kQ}^- ~\eta(p-k)  
\nonumber\\ 
 &+& \frac{3J}{2} ( S_{pQ}^+ - S_{-p-Q~Q}^+ ) \sum_p u_k v_k \eta(p-k)  
  - 4 J~ u_p v_p \sum_k S_{kQ}^+
\label{pi-eom}
\end{eqnarray}
 Here  $\eta(p) = cos p_x + cos p_y $ and the bare dispersion is
renormalized into 
$\tilde{\epsilon}_{p}=\epsilon_{p}-\frac{3J}{2} \sum_k v_k^2~ \eta(p-k) $. 
 The latter corresponds to a trivial rescaling of $t$ which we will 
disregard.

The operator of the collective excitation in the $\pi^+$ channel is 
$\pi_Q^+ = \sum_p g(p) \pi_{pQ}^{\dagger} $. 
 Then from eq.(\ref{pi-d-eom}), we have
\begin{eqnarray}
\left[ {\cal H}, \pi_{Q}^+ \right] = \omega_0
\pi_{Q}^+ + \frac{J \Delta_0}{V_{BCS}} S_Q^+
\label{pi-d}
\end{eqnarray}
where $S_Q^+=\sum_p S_{pQ}^{+}$ and $V_{BCS}$ has been defined earlier.  
Analogously, 
\begin{eqnarray}
\left[ {\cal H}, \pi_{Q}^- \right] = - \omega_0
\pi_{Q}^- - \frac{J \Delta_0}{V_{BCS}} S_Q^+
\label{pi}
\end{eqnarray}
 We can see that in the SC state the EOM for $\pi_Q^{\pm}$ no 
longer close on themselves. 
 The third and the forth terms in eqs.(\ref{pi-d-eom}) and (\ref{pi-eom}), 
that come from anomalous self-energy and scattering correspondingly, 
do not cancel each other exactly. 
 This may be contrasted to the $\eta$-excitation in the negative-$U$ 
Hubbard model\cite{eta}, where exact cancelation of such terms occurs. 

The collective mode in the S channel may be obtained by summing
eq.(\ref{s-eom}) over different $p$'s. 
\begin{eqnarray}
\left[ {\cal H}, S^{+}_{Q} \right] = - \frac{ 2J \Delta_0}{V_{BCS}} \left(
\pi_Q^+ - \pi_Q^- \right)
\label{S_Q}
\end{eqnarray}
 In order to derive this result, we had to disregarded the first term of 
eq.(\ref{s-eom}). 
 In the language of our earlier SCLR approach, this means neglecting 
$\chi_0$ in comparison with $\Delta \chi$. 
 In a close vicinity of $\omega_0$, this is a justifiable assumption, 
because at these frequencies $\Delta \chi$ is strongly peaked and is the 
dominant part of $\chi_{irr}$. 
 However, it is less so at other frequencies, where the incoherent continuum 
is more important. 
 Thus the meaning of going from eq.(\ref{s-eom}) to eq.(\ref{S_Q}) is a 
single mode approximation, which captures collective degrees of freedom only. 
 How good is this approximation? 
 One can see from the numerical results of Sec.\ref{numerix} that for 
$\Delta_0$ around $0.1~J$ the $\pi$-peak already became a dominant feature 
of the $S_Q^+$ spectrum.  
 Some estimates of the realistic value of $\Delta_0$ find it to be close to 
$0.2~J$. 
 In this case, such single mode approximation will truly be a good one 
\footnote{ However, this approximation seems to overemphasize the
  importance of the anomalous scattering terms in the
  energy of the $\pi$ excitation. 
   The anomalous
  self-energy tends to increase the energy of the resonance,
  whereas the anomalous scattering decreases it (see
  eqs.(\ref{pi-d-eom}), (\ref{s-eom}), and (\ref{pi-eom}), for
  example). 
   In Secs.\ref{numerix} and \ref{analyt}, we saw that in the complete
  calculations, the resonance energy in the SC state
  turns out to be above its value in the normal state. 
   However, from eqs.(\ref{pi-d}), (\ref{pi}) and (\ref{S_Q}), we find
  the resonance energy in the SC state to be decreased
  \begin{eqnarray}
    \omega_s^2 = \omega_0^2 - \left( \frac{ 2J \Delta_0 }{ V_{BCS} }
    \right)^2  .
  \end{eqnarray} 
}.

It is instructive to compare our microscopic EOM's with the SO(5) EOM's 
in the SC state\cite{so5}. 
 We have
\begin{eqnarray}
-i \dot{\pi}_{\alpha}^{\pm} &=&  \pm \left( B_{15} \pi_{\alpha}^{\pm} +
  g \langle n_5 \rangle n_{\alpha} \right) 
\label{L-dot}
\\
- i \dot{n}_{\alpha} &=& - \frac{1}{2 \chi_{1\alpha} }\langle n_5
\rangle \left( 
\pi_{\alpha}^{\dagger} - \pi_{\alpha} \right)
\label{n-dot}
\end{eqnarray} 
with $\pi_{\alpha}^{\pm} = L_{1\alpha} \pm i L_{5\alpha} $ and we
assumed dSC ordering along $n_5$. 
 Equation (\ref{L-dot}) is the analogue of eqs.(\ref{pi-d}) and
(\ref{pi}), and eq.(\ref{n-dot}) corresponds to eq.(\ref{S_Q}). 
 Results in this section therefore give a microscopic justification and
quantitative derivation of the $SO(5)$ quantum nonlinear $\sigma$-model
as a long-wave-length theory in the dSC state\cite{so5}.

\section{Summary}
\label{summary}

We have presented detailed analytical and numerical calculations 
for the contribution from the $\pi$ resonance to the spin correlation 
function in the dSC state. 
 The results of these calculations support our earlier interpretation 
of the resonant neutron scattering peak in terms of the p-p collective 
mode in the $\pi$ channel. 
 Various approximations were used in the calculations presented in this work, 
some of them are model-dependent and may not be well-controlled. 
 Therefore, it is important to summarize here the main points leading to 
our conclusion.

1) From general model-independent sum rules on various correlation
functions, one can conclude that the contribution from the
$\pi$ correlation function to the dynamic spin correlation 
function is of the order of $|\Delta|^2/(1-n)$, in excellent
agreement with the two key experimental observations, namely
the vanishing of the sharp mode above $T_c$ and the doping 
dependence of its intensity.

2) Within model-dependent calculations, there is a well-defined 
$\pi$ mode in the p-p channel in the normal state, 
and this mode couples to the p-h spin channel in the dSC
state, where it remains as a sharp excitation. 
 The energy of this mode is not directly related to the dSC gap,
but is directly related to the doping $x$. 
 In the underdoped materials, the dSC gap increases slightly as doping 
is reduced, while the neutron resonance peak energy decreases with $x$. 
 This important experimental finding shows that the neutron resonance
peak is not simply a ``$2\Delta$" phenomenon, and our interpretation
in terms of the $\pi$ resonance naturally resolves this apparent paradox. 
 The doping dependence of both energy and intensity of the neutron
resonance peak were predicted\cite{PRL,so5} before the experiments in
the underdoped superconductors were carried out\cite{fong2,dai1}.

3) Many approximations within our current calculations are not
completely controlled. 
 However, the main behavior can be verified in the case where exact 
knowledge is available. 
 First of all, detailed exact diagonalization studies have been carried 
out both for the $t$-$J$ and the Hubbard models\cite{eder,meixner}. 
 It is clearly seen that the $\pi$ mode in the p-p channel exists in all 
doping range, and it has a low-energy peak where both the energy and 
intensity scale with $x$, in agreement with our$T$-matrix calculation. 
 In contrast to the $\pi$ correlation function, the spin correlation 
function does not have sharp peaks in the high doping range. 
 In the doping range where there are dSC fluctuations, the $\pi$ peak 
coincides with the spin peak.
 From these results, one can conclude that the $\pi$ mode is a genuine
collective mode.
 We can also compare our approximations with the exact $SO(5)$ 
models\cite{silvio}, where the $\pi$ operators are exact eigenoperators of the
Hamiltonian. 
 The manipulations presented in this work lead to results consistent with 
the exact $SO(5)$ Ward identities. 
 The $\pi$ resonance is an exact excitation of the $SO(5)$ models,
and it has exactly the same doping dependence of the mode energy
and intensity as obtained here. 

4) The distinction between the ``RPA peak" and the ``$\pi$ peak"
in the dSC state will be a model-dependent one. 
 In the dSC state, they share the same quantum numbers, and both are based 
on approximate calculations. 
 The origin of the RPA peak may be related to the overestimate of the 
magnetic instability and we see that it may not be robust against 
variations of the vertex corrections or variations of the gap, both of 
which diminish AF instability. 
 On the other hand, the SCLR treatment of the $\pi$ peak
is more robust against these variations. 
 One can test these two approximate schemes within the exact $SO(5)$ models. 
 Only SCLR treatment including the ``$\pi$" process agree with the
exact answer in this case. 
 Therefore, calculations including the ``$\pi$" process is a better 
approximation than the simple RPA calculation.

5) Within our approximations, the spin spectrum consists of an incommensurate
structure at low frequencies, a sharp commensurate peak arising from the 
triplet excitation in the p-p channel (the $\pi$ peak), and  a missing
spectral weight at commensurate wavevector at higher frequencies. 
 These features are in overall agreement with experiments. 
 The predicted weight of the $\pi$ resonance agrees quantitatively with
experiments. 

Therefore, while each of the above arguments are not complete on
their own, the combination of them makes a strong overall case.
 From the interpretation of the neutron resonance peak in terms of the 
$\pi$ mode, we hope to learn a general principle, rather than a specific 
model for fitting a specific experiment. 
 In strongly correlated systems, most degrees of freedoms are strongly coupled,
and most spectra are incoherent. 
 Usually, only a symmetry principle can forbid the decay of a collective 
excitation. 
 In the case of the resonant neutron scattering peak, we believe that it is 
the $SO(5)$ symmetry principle at work, and the $\pi$ mode is the
pseudo Goldstone boson associated with this spontaneous symmetry
breaking. 
 In this paper, we have shown that such an interpretation is consistent 
with the key experimental facts, but it may not be the only possible 
interpretation. 
 Its utility lies in the simplicity and generality of the principle,
which can be applied to other related experiments and lead to new
experimental predictions.

 We would like to express our deep gratitude to Professor
D.J.~Scalapino for his numerous suggestions and constant support
during the course of this work.  
 We would also like to gratefully acknowledge useful conversations 
with J. Brinckmann, R. Eder, H.F. Fong, H. Fukuyama, W. Hanke, 
B. Keimer, P. Lee, S. Meixner, H. Mook and B. Normand.  
 This work is supported by the NSF under grant numbers
DMR-9400372 and DMR-9522915.

\appendix
\section{ Model-independent estimate of the $\pi$
  contribution to the spin susceptibility}

\label{appendix}

A  spectral function for any two operators $A$ and
$B$ is defined as
\begin{eqnarray}
\rho_{A,B}(\omega ) &\equiv& 
                  \frac{1}{2 \pi i} \left[ D^{ret}_{A,B}(\omega) -
                  D^{adv}_{A,B}(\omega) \right] \nonumber\\
 &=&  \sum_n \Bigl[ \langle 0 | A | n \rangle \langle n | B | 0 \rangle 
           \delta ( \omega + E_0 - E_n )     \nonumber \\
 &{}& \hskip 1cm -  \langle 0 | B | n \rangle \langle n | A | 0 \rangle 
           \delta ( \omega - E_0 + E_n )  \Bigr] ,
\label{D1} 
\end{eqnarray} 
where $D^{ret}_{A,B}(\omega)$ and $D^{adv}_{A,B}(\omega)$ are retarded 
and advanced response functions, respectively, 
$| n \rangle$'s are eigenstates of the system Hamiltonian 
with energy $E_n$, and 
$| 0 \rangle$ is the ground state with energy $E_0$. 
If we restrict the above summation only to intermediate states that have  
nonzero overlap with $\pi_\alpha^\dagger | 0 \rangle $, then such
a quantity 
\begin{eqnarray}
\rho^{\pi_\alpha}_{A,B}(\omega )
 &=&  \sum_{n: \langle 0 | \pi_\alpha | n \rangle \neq 0} \Bigl[
    \langle 0 | A | n \rangle \langle n | B | 0 \rangle 
           \delta ( \omega + E_0 - E_n )    \nonumber\\
 &{}& \hskip 2cm - \langle 0 | B | n \rangle \langle n | A | 0 \rangle 
           \delta ( \omega - E_0 + E_n )  \Bigr]        
\label{D3} 
\end{eqnarray}
may be regarded as the contribution of the $\pi$
excitation to the full spectrum $\rho_{A,B}(\omega ) $.  
We can introduce $\omega$-integrated spectral weight as  
\begin{eqnarray}
\bigl[ \rho^{\pi_\alpha}_{A,B} \bigr]_{\omega_1}^{\omega_2} 
 &=& \int_{\omega_1}^{\omega_2} d\omega \rho^{\pi_\alpha}_{A,B}(\omega ) . 
\label{D4} 
\end{eqnarray} 

All of the above spectral functions (or spectral weight)
are bilinear with respect to~$A$ and $B$, and have a property,  
$ \rho_{A,A^\dagger} \geq 0 $ for $\omega \geq 0$ or 
$\omega_2 \geq \omega_1 \geq 0$. 
Therefore, the Cauchy-Schwarz inequality holds provided that the same
frequency condition is satisfied\footnote{ 
This may be proved as follows. From bilinearity and (semi-) positivity, 
we have
\begin{eqnarray}
 0 &\leq& \rho_{A + \lambda B,A^\dagger + \lambda^* B^\dagger} \nonumber \\
 &=& \rho_{A,A^\dagger} + \lambda \rho_{B,A^\dagger} 
   + \lambda^* \rho_{A,B^\dagger} + | \lambda |^2 \rho_{B,B^\dagger} \nonumber 
\end{eqnarray}
for any complex number $\lambda$.  
 Defining $\theta$ as the phase of the \lq mixed correlation function' as 
$\rho_{A,B^\dagger} =  \rho_{B,A^\dagger}^* = |\rho_{A,B^\dagger}| e^{i\theta}$, 
and choosing $\lambda = x e^{i\theta}$ $(-\infty < x < \infty)$, we have 
\begin{eqnarray}
 \rho_{A,A^\dagger} + 2x |\rho_{A,B^\dagger}| + x^2 \rho_{B,B^\dagger}   
 \geq 0 \nonumber
\end{eqnarray}
 Since this inequality holds for any real number $x$, the  
inequality (\ref{C-S}) should hold.}
: 
\begin{eqnarray}
 \rho_{A,A^\dagger} \rho_{B,B^\dagger} \geq | \rho_{A,B^\dagger }|^2 
\label{C-S}
\end{eqnarray}
 Here $\rho_{A,B^\dagger }$ can be either of 
$\rho_{A,B^\dagger }(\omega )$, 
$\rho_{A,B^\dagger }^{\pi_\alpha}(\omega )$ or  
$\bigl[ \rho^{\pi_\alpha}_{A,B^\dagger } \bigr]_{\omega_1}^{\omega_2} $. 

In Section \ref{intro} we derived two sum rules
$\int d\omega \rho^{\pi_\alpha}_{\pi_\alpha,\pi_\alpha^\dagger}(\omega) 
 = 1-n $ and 
$\int d\omega \rho^{\pi_\alpha}_{\pi_\alpha,N_\alpha^\dagger}(\omega) 
 = i\Delta$.
 If most of the $\pi$ spectrum is accommodated in an interval  
$(\omega_0 - \nu, \omega_0 + \nu')$ 
on the positive real axis and 
around the $\pi$-resonance energy $\omega_0$, we can write  
\begin{eqnarray}
 \Bigl[ \rho^{\pi_\alpha}_{\pi_\alpha,\pi_\alpha^\dagger}
 \Bigr]_{\omega_0 - \nu}^{\omega_0 + \nu'}
 \sim 1-n ,  \hskip 2cm 
 \Bigl[ \rho^{\pi_\alpha}_{\pi_\alpha,N_\alpha^\dagger}
 \Bigr]_{\omega_0 - \nu}^{\omega_0 + \nu'}
 \sim i \Delta . 
\end{eqnarray}
 Equation (\ref{C-S}) then immediately gives us
\begin{eqnarray}
 \Bigl[ \rho^{\pi_\alpha}_{N_\alpha ,N_\alpha^\dagger}  
 \Bigr]_{\omega_0 - \nu}^{\omega_0 + \nu'}
 \gsim \frac{|\Delta|^2 }{1-n} .
\label{lower-b}
\end{eqnarray}
 The left hand side of this equation represents the 
contribution of the $\pi$ mode to the spin excitation spectrum 
(e.g., Im$\chi^{zz}/\pi$), and 
the right hand side gives its lower bound \footnote{The (near)
  equality holds if and only if there exists such $ \lambda $ that  
 $ \langle 0 | \pi_\alpha + \lambda N_\alpha | n \rangle = 0 $ 
 for all eigenstates $| n \rangle $ satisfying 
 $ \langle 0 | \pi_\alpha | n \rangle \neq 0$. 
  For example, exact equality holds when $\pi$ operator is an 
 exact eigenoperator 
 and hence there is only one energy eigenstate which satisfies 
 $ \langle 0 | \pi_\alpha | n \rangle \neq 0$.   }. 
  This is a model-independent result. 

 Noting that Im$\chi^{+-} = 2\,$Im$\chi^{zz}$, we obtain eq.(\ref{lower-b1}).
 When applying this result to the present analysis of the $t$-$J$ model, 
where $\Delta_0 = V_{BCS} \Delta = \frac{3J}{2} \Delta$, 
we have  
$ I_\pi \gsim \frac{8}{9J^2} \frac{|\Delta_0|^2}{1-n} $.

 Analysis in this Appendix can be generalized to finite temperatures 
by considering the spectral function,
\begin{eqnarray}
 \rho_{A,B}(\omega) &=&
 \frac{1}{Z} \sum_{n,m} ( e^{-\beta E_n} - e^{-\beta E_m} )
           \langle n | A | m \rangle \langle m | B | n \rangle
          \delta ( \omega + E_n - E_m )  .             
\end{eqnarray}

\end{document}